\documentclass[a4paper,12pt,useAMS]{mn2e}

\usepackage{graphicx}

\title[Independent lensing analyses of Abell 1703]{Full Lensing Analysis of Abell 1703: Comparison of Independent Lens-Modelling Techniques}
\author[Zitrin et al.]{Adi Zitrin$^{1}$\thanks{E-mail:
adiz@wise.tau.ac.il}, Tom Broadhurst$^{1,2,3}$, Keiichi Umetsu$^{4}$, Yoel Rephaeli$^{1}$, Elinor Medezinski$^{1}$ \and ~~~~~~~~Larry Bradley$^{5}$, Yolanda Jim\'enez-Teja$^{6}$, Narciso Ben\'itez$^{6}$, Holland Ford$^{7}$  \and ~~~~~~~~~~~~~~~~~~~~~~~~~~~~~~~~~~~~~~~~~~~~~~~~~~~\emph{and} \and  ~~~~~~~~Jori Liesenborgs$^{8}$, Sven De Rijcke$^{9}$, Herwig Dejonghe$^{9}$, Philippe Bekaert$^{8}$\\\\\\
$^{1}$The School of Physics and Astronomy, the Raymond and Beverly Sackler Faculty of Exact Sciences, Tel Aviv University,\\ Tel Aviv 69978, Israel\\
$^{2}$Department of Theoretical Physics, University of Basque Country UPV/EHU, Leioa, Spain\\
$^{3}$IKERBASQUE, Basque Foundation for Science, 48011, Bilbao, Spain\\
$^{4}$Institute of Astronomy and Astrophysics, Academia Sinica, P.~O. Box 23-141, Taipei 10617, Taiwan\\
$^{5}$Space Telescope Science Institute, 3700 San Martin Drive, Baltimore, MD 21218\\
$^{6}$Instituto de Astrof\'isica de Andaluc\'ia (CSIC), C/Camino Bajo de Hu\'etor, 24, Granada, 18008, Spain\\
$^{7}$Department of Physics and Astronomy, Johns Hopkins University, 3400 North Charles Street, Baltimore, MD 21218\\
$^{8}$Expertisecentrum voor Digitale Media, Universiteit Hasselt, Wetenschapspark 2, B-3590, Diepenbeek, Belgium\\
$^{9}$Sterrenkundig Observatorium, Universiteit Gent, Krijgslaan 281, S9, B-9000, Gent, Belgium}
\begin{document}


\pagerange{\pageref{firstpage}--\pageref{lastpage}} \pubyear{2010}

\maketitle

\label{firstpage}

\begin{abstract}

The inner mass-profile of the relaxed cluster Abell 1703 is analysed by two
very different strong-lensing techniques applied to deep ACS and WFC3 imaging.  Our parametric
method has the accuracy required to reproduce the many sets of multiple images, based on the assumption
that mass approximately traces light. We test this assumption with a
fully non-parametric, adaptive grid method, with no knowledge of the
galaxy distribution. Differences between the methods are seen on
fine scales due to member galaxies which must be included in
models designed to search for lensed images, but on the larger scale the
general distribution of dark matter is in good agreement, with very
similar radial mass profiles. We add undiluted weak-lensing
measurements from deep multi-colour Subaru imaging to obtain a fully
model-independent mass profile out to the virial radius and
beyond. Consistency is found in the region of overlap between the weak
and strong lensing, and the full mass profile is well-described by an
NFW model of a concentration parameter, $c_{\rm vir}\simeq
7.15\pm0.5$ (and $M_{vir}\simeq 1.22\pm0.15 \times
10^{15}M_{\odot}/h$). Abell 1703 lies above the standard
$c$--$M$ relation predicted for the standard $\Lambda$CDM model,
similar to other massive relaxed clusters with accurately determined lensing-based profiles.

\end{abstract}


\section{Introduction}

Simulated CDM dominated halos consistently predict mass profiles that
steepen with radius, providing a distinctive, fundamental prediction
for this form of dark matter (DM; Navarro, Frenk \& White 1996; NFW). Furthermore,
the degree of mass concentration should decline with increasing
cluster mass because in the hierarchical model massive clusters
collapse later, when the cosmological background density is
lower. These predictions are now being subject to stringent lensing
based analyses, using multiply-lensed images and with weak lensing (WL)
information. To date only a few clusters have been reliably analysed
by combining both weak and strong lensing for a full determination of
the mass profile and a definitive comparison with predictions,
(e.g., Gavazzi et al. 2003, Broadhurst et al. 2005b, 2008, Umetsu et al. 2010,
Newman et al. 2009, Okabe et al. 2010 and references therein). The upcoming multi-cycle HST
program of cluster imaging (the \emph{CLASH} program$^{1}$\footnotetext[1]{PI: Postman; http://www.stsci.edu/$\sim$postman/CLASH/}) will provide a much
more definitive derivation of mass profiles for a statistical sample
of relaxed, X-ray selected clusters, combining high
resolution space imaging with deep, wide-field ground based data.

Strong gravitational lensing (SL) is of great significance as a
cosmological probe, providing model-free masses of galaxies and clusters
interior to the Einstein radius and useful constraints
on their inner mass profiles. The mass density in the central regions
of distant clusters typically exceeds the critical value
required for lensing, generating multiple-images of background
objects (e.g., Horesh et al. 2010, Kausch et al. 2010). Recent analyses have shown that many sets of multiply-lensed
images can be uncovered with high-quality space imaging and thanks to
improved modelling techniques. Reliable mass maps are claimed for
several well studied clusters with deep space imaging (e.g., Abell 370, Kneib et al. 1993,
Richard et al. 2010; Abell 611 Newman et al. 2009; Abell 901, Deb et al. 2010; Abell 1689, Broadhurst et al. 2005a, Coe et al. 2010; Cl0024+1654, Liesenborgs et al. 2008, Zitrin et
al. 2009b; MS 2137.3-2353, Gavazzi et al. 2003, Merten et al. 2009; RXJ1347, Brada\v{c} et
al. 2008, Halkola et al. 2008; SDSS J1004+4112, Sharon et al. 2005;
"The bullet cluster", Brada\v{c} et al. 2006).

It is important to realise that most
published mass maps usually adopt an initial model gradient for the cluster mass
profile, rather than deducing and constraining it directly from the data. Various SL modelling methods have developed over the past
two decades in response to the huge improvements in astronomical
imaging. Most methods can be classified as ``parametric" if based on
model prescriptions, or ``non-parametric" if ``grid-based" or
interpolative, capable of arbitrary forms (e.g., Saha \& Williams
1997, Abdelsalam, Saha, \& Williams 1998, Diego et al. 2005,
Liesenborgs et al. 2006, Valls-Gabaud et al. 2006; see also \S 4.4 in
Coe et al. 2008). The non-parametric grid methods do not have the
resolution to accurately locate and reproduce multiple images, and usually rely on images
identified by other means - often just eyeball candidates or
those identified from the subset of parametric models with predictive
power to locate images, for which the number of free parameters does
not exceed the number of independent multiple-images used as
constraints.

The method developed by Broadhurst et al. (2005a), and simplified
further by Zitrin et al. (2009b) has securely identified tens of
multiple images in high quality ACS images, behind Abell 1689 and Cl0024
and also a sample of 12 MACS clusters at $z>0.5$ sample (Zitrin et al. 2010), with only
six free parameters so that in practice the number of
multiple images uncovered readily exceeds the number of free parameters, as minimally required in order to get a reliable fit. This approach to SL is based on the assumption that mass approximately traces light, and will be
employed here as our parametric analysis of Abell 1703. We also apply the
non-parametric technique of Liesenborgs et al. (2006, 2007, 2009) which employs an
adaptive grid inversion technique and has been well tested on available
multiple-imaging data such at the many sets of multiple-images uncovered
by Zitrin et al. (2009b; with photometric redshifts calculated therein), in Cl0024+1654 (see also Liesenborgs et al. 2008).

Here we compare in more detail these two very different methods
applied to Abell 1703, allowing in principle a test of the assumptions
behind the parametric technique. Abell 1703 has been subject to various
complementary studies ranging from early X-ray and optical work to
more recent high quality lensing analyses from ground and space
(Kowalski et al. 1984, B\"{o}hringer et al. 2000, Leir \& Van Den
Bergh 1977, Bade et al. 1998, Cooray et al. 1998, Koester et al. 2007,
Scott et al. 2007, Riley et al. 2008,2009, Bruursema et al. 2008,
Oguri et al. 2009), with detailed radio sources (Rizza et al. 2003,
Coble et al. 2007) and highly magnified high-$z$ galaxies (Zheng et
al. 2009). SL analyses of Abell 1703 have been carried
out by Hennawi et al. (2008), Limousin et al. (2008), de Xivry \& Marshall
(2009), Oguri et al. (2009), Richard et al. (2009), and Saha \& Read (2009).

The cluster Abell 1703 (Abell 1958; see also Abell, Corwin, \& Olowin 1989) is known to have many sets of multiple images with
impressive spectroscopic redshift information (Limousin et al. 2008,
Richard et al. 2009; see also Estrada et al. 2007, Hennawi et
al. 2008). Here we take advantage of two independent SL modelling
techniques which are interesting to compare given their very different
approaches.  We then add the accurate WL data from
Broadhurst et al. (2008; see also Medezinski et al. 2010) to complete
the mass profile for comparison with theoretical predictions
out to the virial radius and beyond and to examine the consistency of
the WL and SL derived profiles in the region of overlap.

 The paper is organised as follows: In \S 2 we describe the
 observations. In \S 3 we detail the SL modelling methods and their
 implementation. In \S 4 we report and discuss the results, which are
 then summarised in \S5. Throughout this paper we adopt a concordance
 $\Lambda$CDM cosmology with ($\Omega_{\rm m0}=0.3$, $\Omega_{\Lambda
 0}=0.7$, $h=0.7$). We adopt a redshift of $z=0.28$ for the cluster, equal to that of the prominent central BCG
 galaxy (Allen et al. 1992). With these parameters one arcsecond corresponds
 to a physical scale of 4.25 kpc for this cluster. The
 reference centre of our analysis is fixed at the centre of the BCG:
RA = 13:15:05.24, Dec = +51:49:02.6 (J2000.0).

\section{Observations}

Abell 1703 was observed in November 2004, with the Wide Field Channel
(WFC) of the ACS installed on HST, in the
framework of the ACS Guaranteed Time Observations (GTO; Ford et
al. 2003) which includes deep observations of several massive,
intermediate-redshift galaxy clusters. Integration times of 7050,
5564, 5564, 8494+1340, $5564\times2$ and $8900\times2$ seconds, were
obtained through the F435W, F475W, F555W, F625W, F775W, and F850LP
filters, respectively, and are available in the Hubble Legacy
Archive. Some important aims of the GTO program are determination of
the mass distribution of clusters for testing the standard
cosmological model and to study distant, background lensed galaxies
for which some of the very highest redshift galaxies are known because
of high magnification by massive clusters (Franx et al. 1997, Frye \&
Broadhurst 1998, Frye, Broadhurst \& Ben\'itez 2002, Kneib et
al. 2004, Stark et al. 2007, Bouwens et al. 2009, Bradley et al. 2008,
in Abell 1703: Zheng et al. 2009).

As part of the ACS GTO cluster program, in April 2010 we also
observed Abell 1703 with the near-infrared channel of HST new Wide
Field Camera 3 (WFC3/IR).  The observations consisted of 1 orbit
(2812s) each in the F125W and F160W bands.

Various redshifts have been quoted for Abell 1703 (e.g., Struble \&
 Rood 1987,1999, B\"{o}hringer et al. 2000), corresponding to several
 different cluster galaxies, out of which we adopt that of the prominent BCG galaxy at $z=0.28$ (Allen et al. 1992). This redshift
 was also used in recent SL work on this cluster by
 Limousin et al. (2008) and Richard et al. (2009) who
identified many sets of multiple-images which we incorporate in this
work, as will be detailed below.

\section{Strong Lensing Modelling and Analyses}\label{model}

\subsection{Parametric Method}

We first apply our well tested approach to lens modelling, which has
previously uncovered large numbers of multiply-lensed galaxies in ACS
images of Abell 1689, Cl0024, and 12 high-$z$ MACS clusters (respectively,
Broadhurst et al. 2005a, Zitrin et al. 2009b, Zitrin \& Broadhurst
2009, Zitrin et al. 2009a, 2010). The full details of
this approach can be found in these earlier papers. Briefly, the basic
assumption adopted is that mass approximately traces light, so that
the photometry of the red cluster member galaxies is used as the
starting point for our model. Cluster member galaxies are identified
as lying close to the cluster sequence by the photometry provided in
the Hubble Legacy Archive.

\begin{figure}
 \begin{center}
   \includegraphics[width=85mm,trim=-10mm 0mm 0mm 0mm,clip]{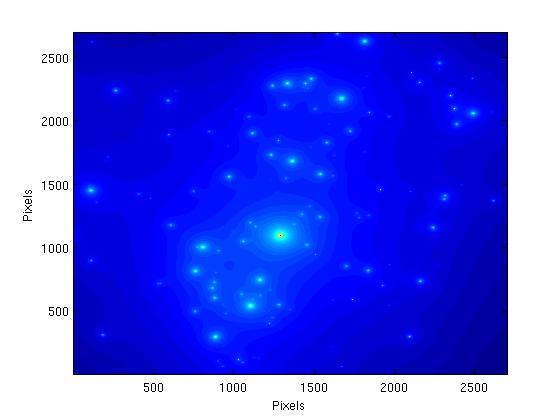}
 \end{center}
\caption{The starting point of the parametric model, where we define the
surface mass distribution based on the cluster member galaxies (see \S \ref{model}) . Axes are in ACS pixels ($0.05 \arcsec /pixel$). North is up, East is left.}
\label{lumpyhcomp}
\end{figure}

\begin{figure}
 \begin{center}
   \includegraphics[width=85mm,trim=-10mm 0mm 0mm 0mm,clip]{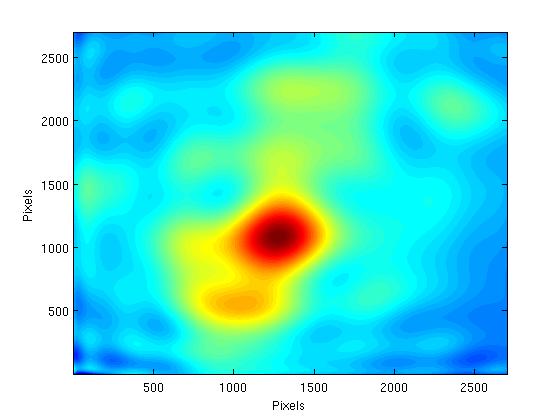}
 \end{center}
\caption{The resulting smooth mass component of the parametric model (see \S \ref{model}). Axes are in ACS pixels ($0.05 \arcsec /pixel$). North is up, East is left.}
\label{smoothcomp}
\end{figure}

 We approximate the large scale distribution of cluster mass by assigning a
 power-law mass profile to each galaxy (see Figure \ref{lumpyhcomp}), the
 sum of which is then smoothed (see Figure \ref{smoothcomp}). The
 degree of smoothing ($S$) and the index of the power-law ($q$) are
 the most important free parameters determining the mass profile. A
 worthwhile improvement in fitting the location of the lensed images
 is generally found by expanding to first order the gravitational
 potential of this smooth component, equivalent to a coherent shear
 describing the overall matter ellipticity, where the direction of the
 shear and its amplitude are free parameters, allowing for some flexibility in
 the relation between the distribution of DM and the distribution of
 galaxies, which cannot be expected to trace each other in detail. The
 total deflection field $\vec\alpha_T(\vec\theta)$, consists of the
 galaxy component, $\vec{\alpha}_{gal}(\vec\theta)$, scaled by a
 factor $K_{gal}$, the cluster DM component
 $\vec\alpha_{DM}(\vec\theta)$, scaled by (1-$K_{gal}$), and the
 external shear component $\vec\alpha_{ex}(\vec\theta)$:

\begin{equation}
\label{defTotAdd}
\vec\alpha_T(\vec\theta)= K_{gal} \vec{\alpha}_{gal}(\vec\theta)
+(1-K_{gal}) \vec\alpha_{DM}(\vec\theta)
+\vec\alpha_{ex}(\vec\theta),
\end{equation}
where the deflection field at position $\vec\theta_m$
due to the external shear,
$\vec{\alpha}_{ex}(\vec\theta_m)=(\alpha_{ex,x},\alpha_{ex,y})$,
is given by:
\begin{equation}
\label{shearsx}
\alpha_{ex,x}(\vec\theta_m)
= |\gamma| \cos(2\phi_{\gamma})\Delta x_m
+ |\gamma| \sin(2\phi_{\gamma})\Delta y_m,
\end{equation}
\begin{equation}
\label{shearsy}
\alpha_{ex,y}(\vec\theta_m)
= |\gamma| \sin(2\phi_{\gamma})\Delta x_m -
  |\gamma| \cos(2\phi_{\gamma})\Delta y_m,
\end{equation}
where $(\Delta x_m,\Delta y_m)$ is the displacement vector of the
position $\vec\theta_m$ with respect to a fiducial reference position,
which we take as the lower-left pixel position $(1,1)$, and
$\phi_{\gamma}$ is the position angle of the spin-2 external
gravitational shear measured anti-clockwise from the $x$-axis.
The normalisation of the model and the relative scaling of the
smooth DM component versus the galaxy contribution brings the
total number of free parameters in the model to 6. This approach to SL is
sufficient to accurately predict the locations and internal
structure of multiple images, since in practice the number of
multiple images uncovered readily exceeds the number of free parameters thus fully constraining them.

\begin{figure*}
 \begin{center}
  \includegraphics[width=130mm,trim=0mm 0mm 0mm 0mm,clip]{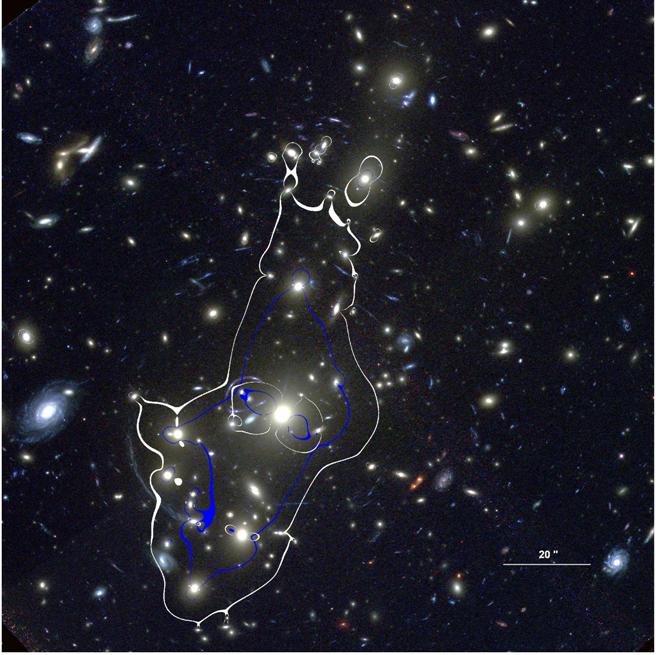}
 \end{center}
\caption{Galaxy cluster Abell 1703 ($z=0.28$) imaged
with Hubble/ACS. The overlaid critical curve (blue) corresponds to the
distance of system 1, at $z_{s}=0.889$. The outer white critical curve corresponds to the giant arc (systems 10/11) at higher redshift, $z_{s}=2.627$, enclosing a critical area of an effective Einstein radius of $\simeq 130$ kpc at the redshift of this
cluster. North is up, East is left.}
\label{curves1703}
\end{figure*}

In addition, two of the six free parameters can be primarily set to reasonable values so only 4 of
these parameters have to be constrained initially, which sets a very
reliable starting-point using obvious systems. The mass distribution
is therefore primarily well constrained, uncovering many
multiple-images which can be then iteratively incorporated into the
model, by using their redshift estimation and location in the
image-plane.

\begin{figure*}
 \begin{center}
  \includegraphics[width=150mm,trim=-4mm 0mm 0mm 0mm,clip]{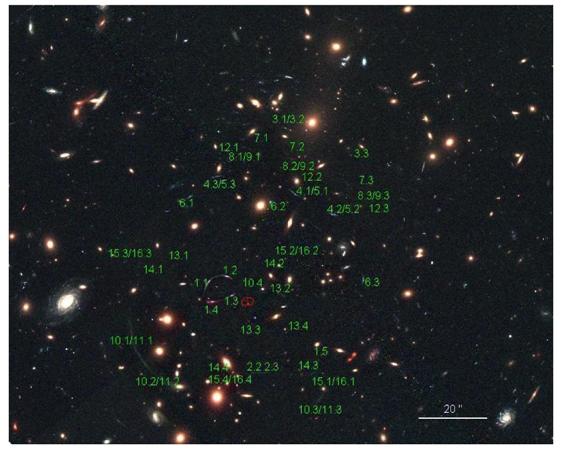}
 \end{center}
\caption{Distribution of multiply-lensed images used to constrain the models marked on a colourful Hubble/ACS image of Abell 1703, with some central galaxies including the cD galaxy subtracted, and its original location marked in red. Subtraction has been carried out by modelling the main cD galaxy and three other galaxies using Chebyshev-Fourier basis functions (``CheF-lets''; Jim\'enez-Teja \& Ben\'itez, in preparation). These multiply-lensed images and more details can be found in Limousin et al. (2008) and Richard et al. (2009). North is up, East is left.}
\label{images1703}
\end{figure*}

\begin{figure}
 \begin{center}
  \includegraphics[width=80mm,trim=0mm 0mm 0mm -5mm,clip]{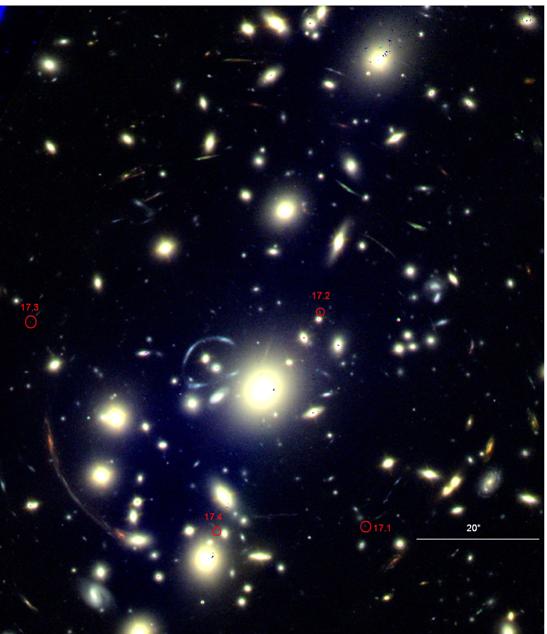}
 \end{center}
\caption{An optical/ACS + IR/WFC3 colourful image of the central region of Abell 1703. The much wider colour range enabled by incorporation of the IR data manifests the distinct colours of the many multiple-systems shown in Figures \ref{curves1703} and \ref{images1703}, and the identification of an additional system (17) marked in circles, following the same symmetry of systems 15 and 16 (Figure \ref{images1703}).}
\label{OPTIR1703}
\end{figure}

Firstly we use this preliminary model to lens the more obvious lensed
galaxies back to the source plane by subtracting the derived
deflection field, and then relens the source plane to predict the
detailed appearance and location of additional counter images, which
may then be identified in the data by morphology, internal structure
and colour. We stress that multiple images found this way must be
accurately reproduced by our model and are not simply eyeball
``candidates'' requiring redshift verification. In Abell 1703 many multiple-images (16 systems, most of them with spectroscopic redshifts; Limousin et al. 2008, Richard et al. 2009) are already known and therefore simply used to constrain the fit, which is assessed by the RMS uncertainty in the image plane:

\begin{equation} \label{RMS}
RMS_{images}^{2}=\sum_{i} ((x_{i}^{'}-x_{i})^2 + (y_{i}^{'}-y_{i})^2) ~/ ~N_{ima
ges},
\end{equation}
where $x_{i}^{'}$ and $y_{i}^{'}$ are the locations given by the
model, and $x_{i}$ and $y_{i}$ are the real images location, and the
sum is over all $N_{images}$ images. The best-fit solution is unique
in this context, and the model uncertainty is determined by the
location of predicted images in the image plane. Importantly, this
image-plane minimisation does not suffer from the well known bias
involved with source plane minimisation, where solutions are biased by
minimal scatter towards shallow mass profiles with correspondingly
higher magnification.

The model is successively refined as additional sets of multiple
images are incorporated to improve the fit, using also their redshift
measurements or estimates, for better constraining the mass slope
through the cosmological relation of the $D_{ls}/D_{s}$ growth.

\subsection{Non-parametric inversion method}

\subsubsection{Genetic algorithm based inversion}

The non-parametric inversion method that we apply here is based on the
work of Liesenborgs et al. (2006). It requires the user to specify a
square-shaped region in which the inversion routine should try to
reconstruct the mass distribution. Additionally, it is necessary to
define which images correspond to the same source and at what
redshifts the sources are located. In a first step, the square region
is subdivided in a uniform way into a number of smaller square grid
cells, and to each cell a projected Plummer sphere (Plummer, 1911) is
assigned. The width of each basis function is set proportional to the
grid cell size. As an additional basis function, a sheet of mass can
be included; this can be useful as in the center of clusters a
non-negligible density offset may be present which can prove difficult
to model using Plummer basis functions.  A genetic algorithm is then
used to search for appropriate weights of these basis functions,
yielding a first approximation of the projected density of the lens.

Using this first approximate solution, a new grid is then constructed
in which regions containing more mass are subdivided further. It
should be noted that the mass sheet basis function is not taken into
consideration in this step as it is structureless. Using this new
grid, basis functions are assigned and the genetic algorithm again
looks for appropriate weights. This refinement procedure can be
repeated until the added resolution no longer results in an improved
fit to the data.

The actual search for appropriate weights of the basis functions and
thus for the mass distribution, employs a genetic algorithm.
This is a heuristic optimisation strategy, inspired by the theory of
evolution by Darwin. In essence, one tries to breed solutions to a
problem, by evolving an initial population of trial solutions towards
solutions which are better adapted to the problem under study. To
create the next generation from the current one, trial solutions are
combined and mutated, while applying selection pressure, i.e., making
sure that solutions which are deemed better create more offspring. In
this approach it is even possible to simultaneously optimise against
several so-called fitness measures; one then speaks of a
multi-objective genetic algorithm (see, e.g., Deb 2001).

As the number of basis functions used can become quite large
(e.g. $\sim$ 1000) and the genetic algorithm starts from random
initial solutions, several runs of this procedure will produce results
that differ somewhat. Therefore a set of solutions is usually
generated, so that one can inspect the common features of these mass
maps, and the standard deviation can be used as a measure of
the reliability at any location. The algorithm details and original
fitness criteria are described in Liesenborgs et al. (2006, 2007,
2009).  Below, the fitness criteria used in this work shall be
described. It is based on these criteria that selection pressure will
be applied.

\subsubsection{Fitness criteria}

In strong lens inversion one tries to deduce the projected mass
distributions based on data of multiply imaged sources. Since each
set of images originates from a single source, projecting the images
back onto the corresponding source plane should produce a consistent
source. Previously, only extended images could be used, in which case
the back-projected images should overlap in the source plane. To
calculate the amount of overlap, the estimated size of the source
was used as a length scale. The method was adapted to work with point
images as well. In this case the envelope of all estimated source
positions is used as a length scale when calculating the distances
between the back-projected images of each source. Using this length
scale instead of an absolute scale, prevents scaling the source plane
to obtain a better fitness value. This is especially important when a mass sheet is included as a basis function since it has precisely
this effect.

Apart from the locations where images can be seen, additional constraints
come from the area in which no images are observed, i.e., the null space.
To avoid predicting extra images, which corresponds to avoiding unnecessary
substructure, the user can define a region which will be used to check for
additional images. This region is divided into a large number of triangles
and for each source, the triangles are projected onto the source plane.
Each triangle that overlaps with the envelope of the back-projected
images is counted and the total count for all sources is used as the
null space fitness measure. As for each source this gives an approximation
of the amount of images, a lower value indicates a better fitness with
respect to this criterion.

\subsubsection{Input}

The input of the inversion routine consists of multiply imaged
systems together with their redshifts. The systems listed in
Richard et al. (2009) were used for this purpose, and where
available, spectroscopic redshift information was used. For systems
2, 8, 9 and 12, the redshift predicted by the model in this
work was used, as these redshifts were in good agreement with the
photometric ones. Note, systems 13 and 14 were not used in the inversion
as the redshift estimates seemed more uncertain (however we do find that including them results only in minor changes to the mass model). Based on these image systems, the inversion routine was instructed to look for mass in a 2$\times$2
arcmin$^2$ region, roughly centered on the cD galaxy.
To limit the amount of predicted images that were not part of the
input, and therefore to limit the amount of unnecessary substructure,
the null-space region was 3$\times$3 arcmin$^2$
in size, and a 48$\times$48 grid was used for calculating
the corresponding fitness measure.

\section{Results and Discussion}

\subsection{Strong-Lensing Regime and cD Galaxy}

In the SL regime we have modelled Abell 1703 (see Figure \ref{curves1703}) using the many sets of
multiply-lensed images previously identified, most of which have
spectroscopic redshifts reported in Limousin (2008) and Richard et
al. (2009). Here we incorporate also new WFC3/IR imaging, which reveals the distinct colours of each system, and enables the identification of an additional system (17) following the same symmetry as systems 15/16 (see Figures \ref{images1703} and \ref{OPTIR1703}) with a similar model redshift of $z_s\sim2.8$. We have used the parametric method of Zitrin et al. (2009b) to further verify the reliability of the many multiply-lensed images across the field and to securely input them into the non-parametric method of Liesenborgs et al. (2006) to model the central
mass distribution.

\begin{figure}
 \begin{center}
   \includegraphics[width=85mm, trim=0mm 0mm 0mm 0mm,clip]{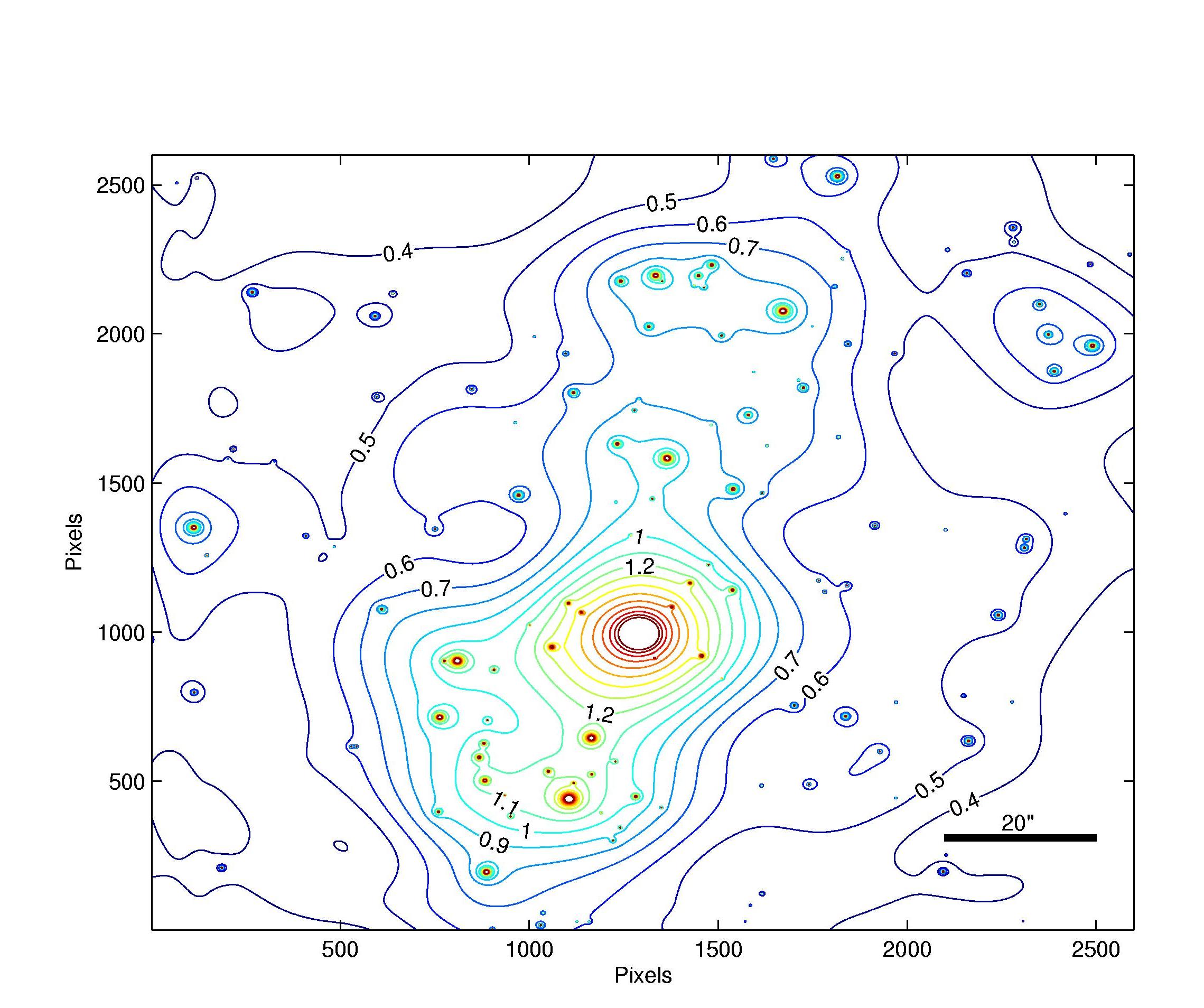}
 \end{center}
\caption{2D surface mass distribution ($\kappa$), in units of the
critical density (for $z_s=2.627$), of Abell 1703. Contours are shown
in linear units, derived from the parametric mass model constrained
using the many sets of multiply-lensed images seen in Figure
\ref{images1703}. Axes are in ACS pixels ($0.05 \arcsec /pixel$), North is up, East is left.}
\label{contoursAdi}
\end{figure}

\begin{figure}
 \begin{center}
   \includegraphics[width=85mm,trim=0mm 0mm 0mm 0mm, clip]{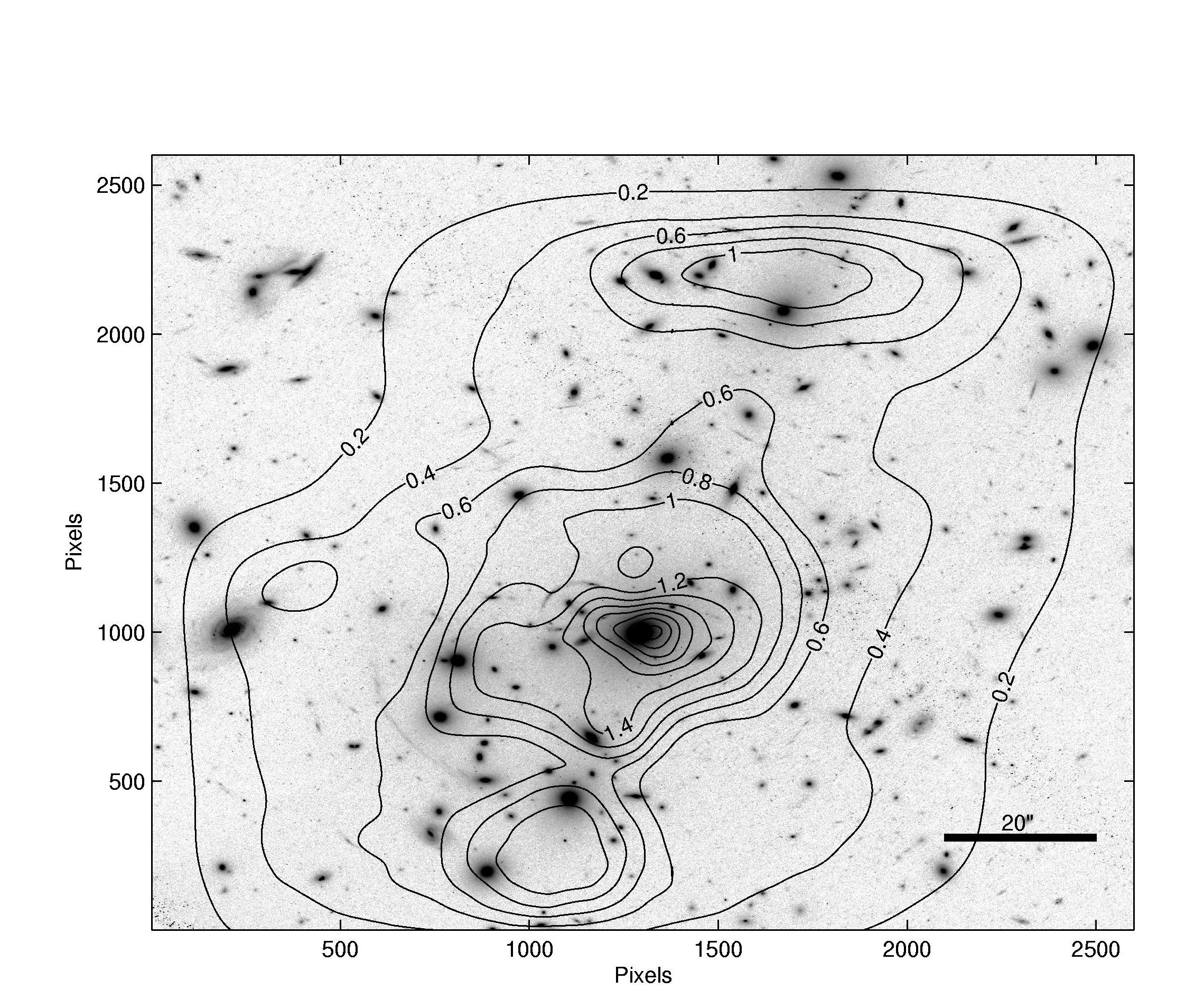}
 \end{center}
\caption{2D surface mass distribution ($\kappa$) contours overlayed on the cluster image,
derived from the non-parametric mass model constrained using the
multiply-lensed images seen in Figure \ref{images1703}. As can be
seen, though no galaxies were included in this modelling method, the
mass contours steepen up where significant galaxies are present. Clearly this
mass distribution is in good agreement with the parametric mass
distribution seen in Figure \ref{contoursAdi}.}
\label{contoursJori}
\end{figure}

Our parametric model (see Figure \ref{contoursAdi}) accurately reproduces all
multiply-lensed images, indicating that our preliminary
assumption that mass traces light is reasonable. In order to further test this assumption we then applied the non-parametric technique of Liesenborgs et al. (2006, 2007, 2009) for which no prior information for the
distribution of cluster galaxies or mass is input. Nevertheless, the
results of this method seem to trace the distribution of light as can
be seen in Figure \ref{contoursJori}, generating a 2D mass distribution
which is remarkably similar to the result of the parametric mass model
on the large scale, shown in Figure \ref{contoursAdi}. A distinct
substructure is seen in both maps and corresponds to local galaxy over-densities. In addition, the two methods produce very similar mass
profiles over a range of scales covering the
full distribution of multiple images, with a mean inner slope of $d\log \Sigma/d\log
\theta\simeq -0.5$.

\begin{figure}
 \begin{center}
   \includegraphics[width=85mm, trim=0mm 0mm 0mm 0mm, clip]{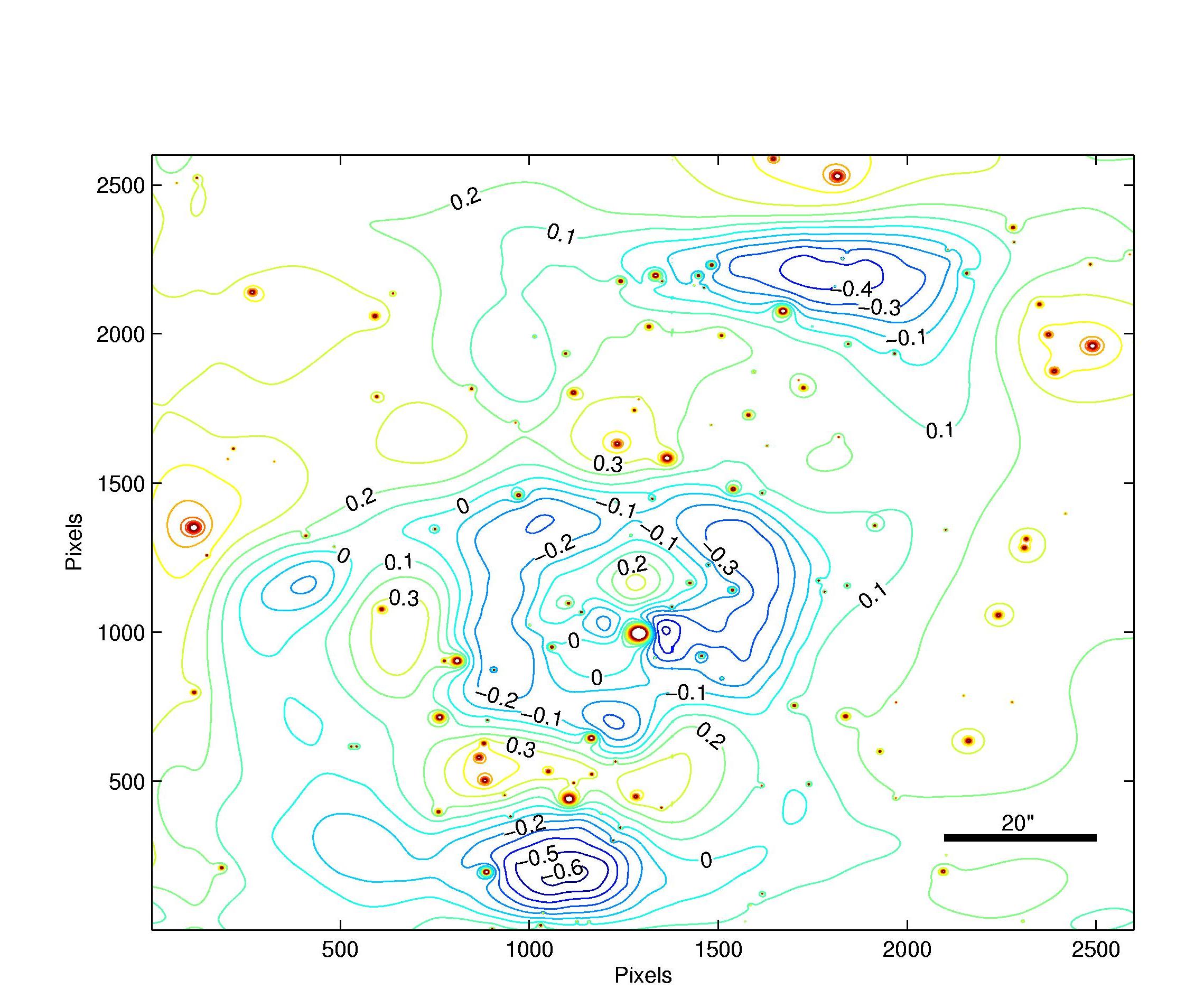}
 \end{center}
\caption{A 2D contour map of the difference between the parametric (Figure \ref{contoursAdi}) and the non-parametric (Figure \ref{contoursJori}) mass distributions. Contours are in $\Delta\kappa$, plotted in equal linear spaces of 0.1. The main positive differences (red) are seen where galaxies are located in the data (since these are included only in the parametric model), and the main negative differences (blue) are seen where the non-parametric model has more mass than implied by the galaxy distribution. Still these differences are overall small, with a mean difference of $|\Delta\kappa|$=0.19 across this field, contributed mainly by the inclusion of cluster members or discrepancies outside the critical curves, where one has relatively poor constraints from the observed multiple images.}
\label{contoursDif}
\end{figure}

We have examined the difference between these two mass maps by subtracting the non-parametric mass distribution from the parametric mass distribution. The result is shown in Figure \ref{contoursDif}. As can be seen in this figure, the main positive differences, marked in red, occur mainly where galaxies are located in the data, since these must contain mass and are included only in the parametric model. The non-parametric model does place mass at these locations, but it is usually smoothly distributed as this approach to modelling does not make prior assumptions about the mass distribution and thus does not achieve a spatial resolution sufficient for resolving individual cluster galaxies. The main negative differences, marked in blue, are seen where the non-parametric model has more mass than implied by the galaxy distribution, but overall these are small and likely inevitable given the inherent noise set by the finite amount of input data. The mean difference across this field is $|\Delta\kappa|=0.19$, contributed mainly by the inclusion of cluster members or discrepancies outside the critical curves, where one has relatively poor constraints from the observed multiple images.

The parametric method of Zitrin et al. (2009b) has been shown to have
the predictive power to find many multiple images in the field. This
parametric method has inherently more structure on small scales by
virtue of the inclusion of cluster members which can significantly
deflect images locally and must be included in order to find lensed
images. Due to the low number of parameters this model is initially
well-constrained using only a few sets of usually obvious
multiple-systems, thus correlated to the initial mass distribution so that the image-plane reproduction accuracy can be only somewhat improved as
newly-found multiple-systems are incorporated, but the overall gradient of the cluster lensing profile is
significantly refined through the cosmological distance-redshift
relation. It is important to have a wide range of background source
redshifts for a reliable profile determination, otherwise the SL
models are degenerate with respect to the profile, although the
relative distribution of matter and substructure can still be reasonable, finding many sets of multiple-images. In the non-parametric approach on the
other hand, the fit is much more flexible and is continuously improved
as more of these images are incorporated and the overall solution is
clarified, allowing the exclusion of a wide range of non-unique
solutions (Liesenborgs et al. 2006, 2008). When sufficient images
are incorporated, the overall mass distributions and profiles of these
two methods become very similar, as we have found here and is shown in the comparison of Figures \ref{contoursAdi}, \ref{contoursJori} and \ref{contoursDif}.

\begin{figure}
 \begin{center}
  \includegraphics[width=60mm,trim=0mm 0mm 0mm 0mm,clip]{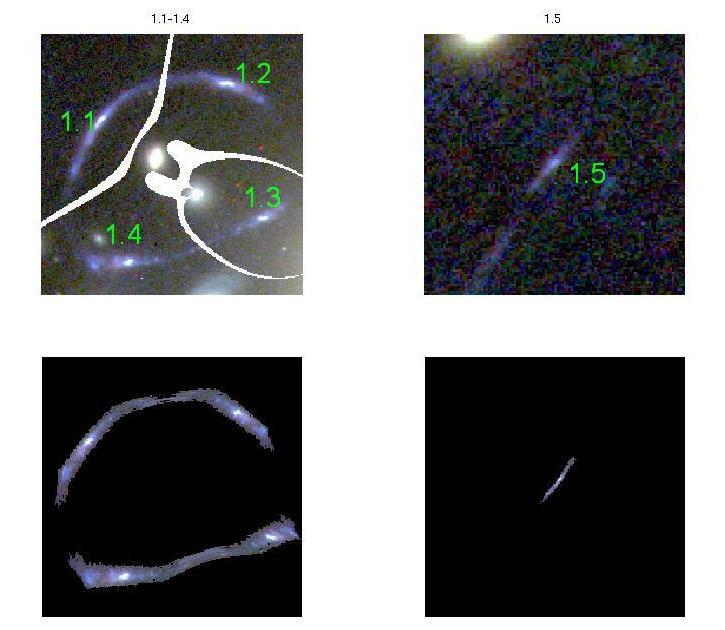}
 \end{center}
\caption{Reproduction of system 1 by our model, by delensing
image 1.4 into the source plane, and then relensing the source plane pixels onto the image plane to accurately form the ring. By tuning
the inner slope of the mass distribution, the observed structure is reproduced very closely. A small fifth image, 1.5, is
formed outside the tangential critical curve (for $z_s=0.889$; see Figures \ref{curves1703} and \ref{images1703}).}
\label{rep_sys1}
\end{figure}

\begin{figure}
 \begin{center}
   \includegraphics[width=85mm,trim=-7mm 0mm 0mm 0mm,clip]{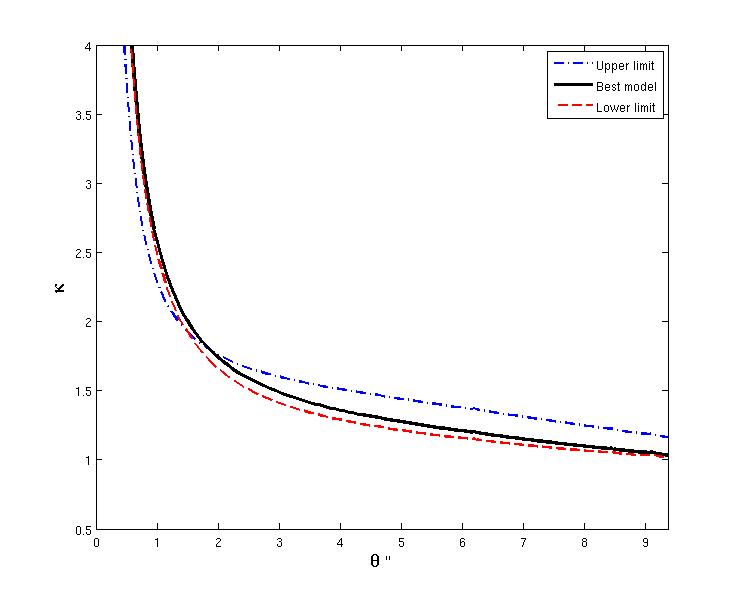}
 \end{center}
\caption{The four images of the ring (system 1, $z_s=0.889$; see
also Figure \ref{rep_sys1}) next to the cluster core enable a
unique determination of
the cD galaxy mass-profile in the range $\simeq1-5\arcsec$ (i.e.,
$\simeq4-22$ kpc; where the closest image of the ring appears). The
solid thick black line represents the best-fit parametric model which
accurately reproduces the ring. The dashed red line represents the
minimum ``shallowness'' threshold, meaning that only models higher
than this threshold will form the four images of the ring, and the
dash-dotted blue line shows the limit above which these images become
overly distorted.}
\label{cdprofile}
\end{figure}

The critical curves for different sources are plotted on the cluster
image in Figure \ref{curves1703}. The critical curves for a source
redshift of $z_s=2.627$ enclose an area with an effective Einstein
radius of $30.5\pm3\arcsec$ ($\simeq$130 kpc at the redshift of the
cluster), and a mass of $1.25\pm0.1 \times 10^{14} M_{\odot}$. Our
parametric model reproduces all multiply-lensed systems within a
$\sigma_i=1.5\arcsec$ of their real location. In particular, our
best-fit parametric model accurately reproduces the complicated ring
($z_s=0.889$; system 1) as can be seen in Figure
\ref{rep_sys1}. We notice that only
models with central mass-profiles steeper than a certain threshold
accurately reproduce all parts of the ring, importantly enabling us to
constrain the mass and the profile of the cD galaxy in the range
$\simeq1-5 \arcsec$, where the closest image of the ring system forms. This is
shown also in Figure \ref{cdprofile}.

We find that the cD galaxy encloses a \emph{projected} mass of $5.2\pm0.4 \times
10^{11} M_{\odot}$ within a radius of $\simeq5\arcsec$ ($\simeq$22 kpc) after
subtracting the interpolated smooth DM component ($\simeq6.3\times
10^{12} M_{\odot}$ inside this aperture), and has a B-band
luminousity of $8.8\pm0.1 \times 10^{10} L_{\odot}$ (fluxes were
converted to luminousities using the LRG template described in
Ben\'itez et al. 2009). This corresponds to an averaged $M/L_B$ of
$\sim6~(M/L)_{\odot}$ in this region. This ratio can be fully
accounted for by the stars contained in this galaxy, for which we
obtain as well $M/L_B \simeq 6~(M/L)_{\odot}$, for a single-burst
stellar population formed at $z = 3$ and viewed at a redshift of $z =
0.28$, equivalent to an age of $\simeq$8.1 Gyrs, and with half solar
metallicity (by evolutionary models of Bruzual \& Charlot 2003). This
result is in agreement with the result of Limousin et al. (2008), and
is similar to other lensing based cD masses in well-studied clusters,
for which low M/L ratios are also found and fully accounted for by the
measured stellar light (e.g., Gavazzi et al. 2003 for MS2137-2353, and Zitrin \&
Broadhurst 2009 for MACS J1149.5+2223).

\begin{figure}
 \begin{center}
   \includegraphics[width=80mm,trim=0mm 0mm 0mm 0mm,clip]{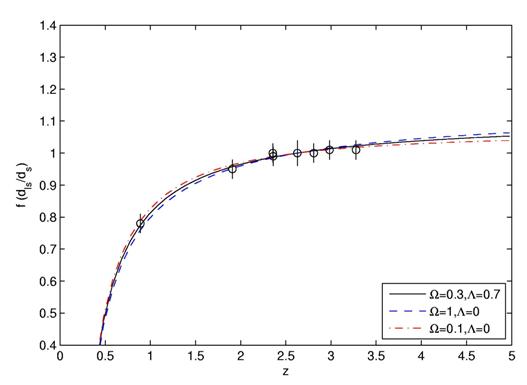}
 \end{center}
\caption{Growth of the scaling factor $f(d_{ls}/d_{s})$ as a function of
redshift, normalised so f=1 at $z=2.627$. Plotted lines are the
expected ratio from the chosen specified cosmological model. The circles
correspond to the multiple-image systems reproduced by the parametric mass model, versus their real spectroscopic redshift. The data follow very well the
relation predicted by the standard cosmological model (mean deviation of only $\Delta_{f}< 0.01$, and $\chi^{2}=0.1$
for this fit).}
\label{dlsds}
\end{figure}

\begin{figure}
 \begin{center}
 \vspace{0.5cm}
  \includegraphics[width=60mm,trim=-3mm 0mm 0mm 0mm,clip]{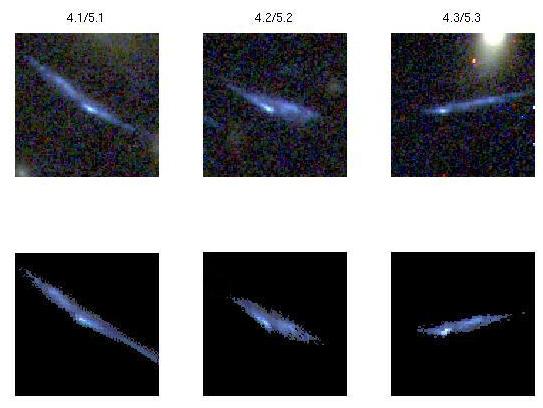}
 \end{center}
\caption{Reproduction of systems 4 \& 5 by our model, by delensing image 4.1/5.1 into the source plane, and relensing the source-plane pixels onto the image plane to accurately form the other images of this system.}
\label{rep_sys45}
\end{figure}

\begin{figure}
 \begin{center}
 \vspace{0.5cm} 
 \includegraphics[width=60mm,trim=-3mm 0mm 0mm 0mm,clip]{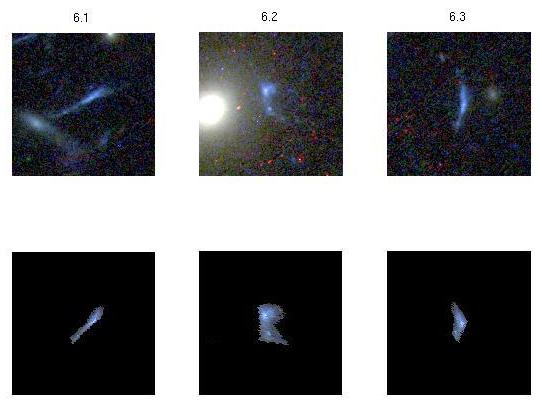}
 \end{center}
\caption{Reproduction of system 6 by our model, by delensing image 6.2 into the source plane, and relensing the source-plane pixels onto the image plane to accurately form the resolved internal details of the other images of this system.}
\label{rep_sys6}
\end{figure}

\begin{figure}
 \begin{center}
 \vspace{0.5cm}
  \includegraphics[width=60mm,trim=-5mm 0mm 0mm 0mm,clip]{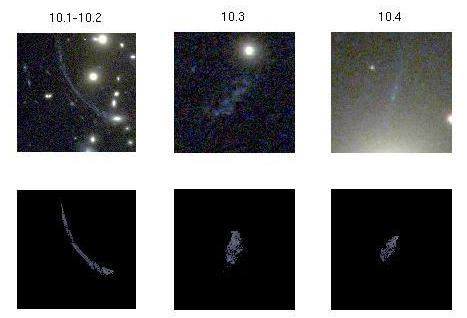}
 \end{center}
\caption{Reproduction of the giant arc, systems 10 \& 11, by our model. We
delens image 10.2 into the source plane, and relens the
source-plane pixels onto the image plane, which accurately reproduces also the other half of the main arc, image 10.1, and two other smaller images (10.3 and 10.4; see Figure \ref{curves1703}).}
\label{rep_sys10}
\end{figure}

We mentioned in the preceding sections that the profile can only be accurately constrained by
incorporating the cosmological redshift-distance relation, i.e., the lensing distance of each system based on the measured spectroscopic redshifts.
In so doing we make use in particular of the $z=0.889$ system (the ring; system number 1),
whose redshift is very distinct from the rest of the multiple-image systems, thus strongly constraining the profile.
We examine how well the cosmological relation is reproduced by the parametric model, accounting also for all other systems with reliable spec-$z$ measurements, as shown in Figure \ref{dlsds}. Clearly the redshifts of these systems verify very well that the predicted deflection of the
best fitting model at the redshift of each of these systems,
lies precisely along the expected cosmological relation, with a mean deviation of only $\Delta_{f}< 0.01$ (see Figure \ref{dlsds}), and $\chi^{2}=0.1$
for the best model, considerably strengthening the plausibility of our parametric approach to modelling in general. In Figs \ref{rep_sys45}, \ref{rep_sys6} and \ref{rep_sys10} we give further examples demonstrating how different systems are accurately reproduced by our parametric model, in addition to the remarkable reproduction of the ring system seen in Fig. \ref{rep_sys1}.

\subsection{Combined Weak and Strong Lensing}

We now compare our SL analysis results with undiluted WL data from deep Subaru $g'r'i'$ images (Broadhurst et al. 2008).
The Subaru WL data, covering a wide field of $\approx 34\arcmin\times
27\arcmin$, allow us to probe the cluster mass distribution over a
wide radial range, $\theta\approx [0.7\arcmin,18\arcmin]$, in the
subcritical regime ($\theta > 30\arcsec$).

\begin{figure}
 \begin{center}
   \includegraphics[width=60mm,angle=270,trim=0mm 0mm 0mm 0mm,clip]{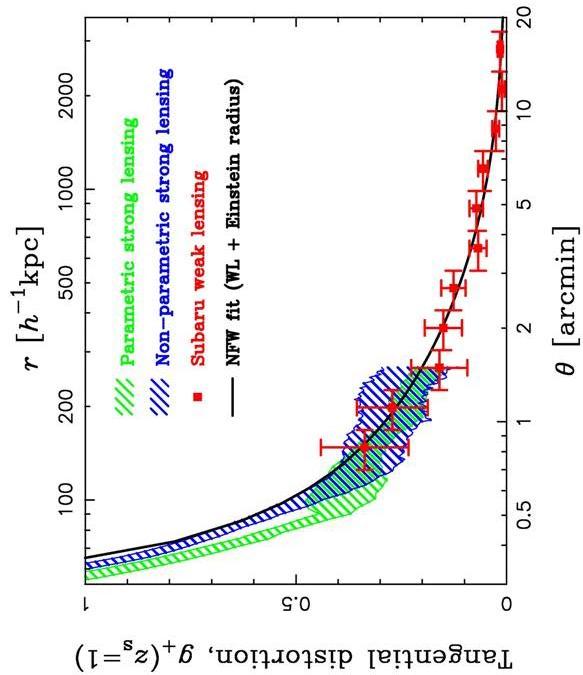}
 \end{center}
\caption{Comparison of
our parametric and non-parametric SL profiles with the Subaru
distortion profile. Our SL and WL results are in good agreement where
the data overlap, $\theta=[40\arcsec,90\arcsec]$, for both SL methods.
Furthermore, a simple inward extrapolation of the best-fitting NFW
profile (black solid curve; see also Table \ref{NFWfitTable}) for the
outer Subaru observations with input of the Einstein radius, fits well
with the inner SL information, in particular, slightly better for the
non-parametric profile. The parametric profile has a minor ``bump''
around $\sim30\arcsec$ due to other bright galaxies in the field which
are not included in the non-parametric model, translating into a dip in the $g_{+}(\theta)$ profile of the parametric technique,
slightly deviating from the smooth NFW curve.}
\label{SWgprofile}
\end{figure}

For a direct comparison with the WL data, we followed the method
outlined in Umetsu et al. (2010) to translate our SL mass profiles
into corresponding tangential distortion profiles
$g_{+}(\theta)=\gamma_{+}(\theta)/(1-\kappa(\theta))$ for a fiducial
source redshift $z_s=1$, roughly matching the mean depth of blue+red
background galaxies used for the WL analysis.  In Figure
\ref{SWgprofile} we compare our parametric and non-parametric SL inner
profiles with the Subaru distortion profile. Our SL and WL results are
in good agreement where the data overlap,
$\theta=[40\arcsec,90\arcsec]$, for both SL methods. Furthermore, a
simple inward extrapolation of the best-fitting NFW profile (see Table
\ref{NFWfitTable}) for the outer Subaru observations with input of the
Einstein radius, fits well with the inner SL information, in
particular, slightly better for the non-parametric profile. The
parametric profile has a minor ``bump'' around $\sim30\arcsec$ due to
other bright galaxies in the field that are not included in the
non-parametric model. This translates into a dip in the
$g_{+}(\theta)$ profile of the parametric technique, slightly
deviating from the smooth NFW curve. Furthermore, the WL+Einstein
radius NFW fitting assumes a circularly symmetric lens, possibly
biasing this NFW profile fit done this way.

\begin{figure}
 \begin{center}
   \includegraphics[width=60mm,angle=270,trim=0mm 0mm 0mm 0mm,clip]{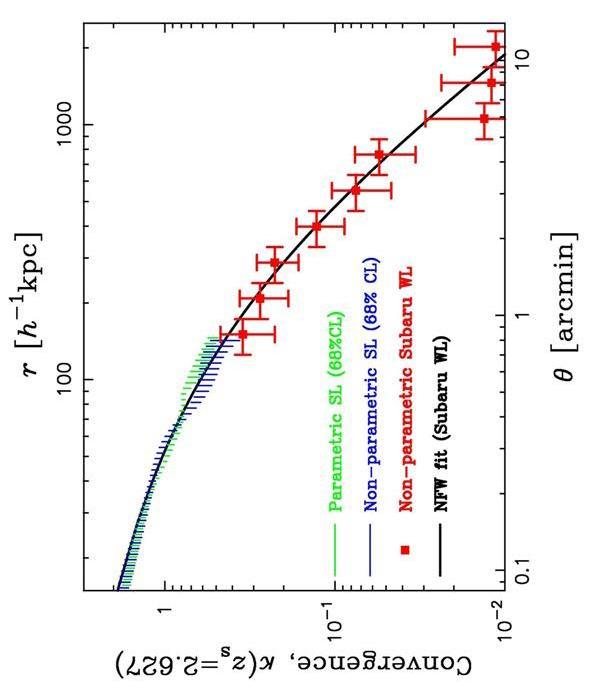}
 \end{center}
\caption{Comparison of our SL and WL results in terms of the
lens convergence profile, $\kappa(\theta)$. The combined SL and WL
results produce a coherent mass profile with a continuously steepening
radial trend from the central region to the outskirts of the
cluster. Both SL methods are in very good agreement with the WL data
and the NFW fit (see also Table \ref{NFWfitTable}).}
\label{SWprofile}
\end{figure}

We then reconstruct the outer mass profile from the Subaru WL data
using the shear-based one-dimensional inversion method outlined in
Umetsu et al. (2009, 2010).  Figure \ref{SWprofile} compares our SL
and WL results in terms of the lens convergence profile,
$\kappa(\theta)$, where the combined SL and WL results produce a
coherent mass profile with a continuously steepening radial trend from
the central region to the outskirts of the cluster.

Overall, we combine our SL and WL results to examine the form of the
underlying cluster mass profile and to characterise cluster mass and
structure properties. To do this, we fit our SL and WL constraints
with an NFW profile in four independent manners: First, we fit the
Subaru distortion profile $g_+(\theta)$ alone with no SL information
involved. Second, we utilise the inner Einstein radius constraint,
$\theta_{\rm E}=30.5\pm 3\arcsec$ ($10\%$ uncertainty) at $z_s=2.627$,
in conjunction with the Subaru distortion profile (for details, see
Umetsu \& Broadhurst 2008 and Umetsu et al. 2010).  Most credibly, we
then fit our joint SL+WL convergence profiles $\kappa(\theta)$ of the
parametric and non-parametric SL methods following the prescription
given by Umetsu et al. (2010). For the joint NFW fitting we constrain
the SL data to the range [5\arcsec, 25\arcsec] where the two
independent SL profiles are extremely similar, and since in general
the shapes of the SL mass profiles within the Einstein radius look smoother,
matching better the expectation from the outer WL profile.  All these
methods yield similar and consistent results, which are also
summarised in Table
\ref{NFWfitTable}.

The NFW universal density profile has a two-parameter
functional form as (Navarro, Frenk, \& White 1997)
\begin{eqnarray}
\label{eq:NFW}
 \rho_{\rm NFW}(r)= \frac{\rho_s}{(r/r_s)(1+r/r_s)^2},
\end{eqnarray}
where $\rho_s$ is a characteristic inner density, and $r_s$ is a
characteristic inner radius.
The logarithmic gradient $n\equiv d\ln\rho(r)/d\ln r$
of the NFW density profile flattens
continuously towards the center of mass, with a flatter central slope
$n=-1$ and a steeper outer slope ($n\to -3$ when $r\to \infty$)
than a purely isothermal structure ($n=-2$).
A useful index, the concentration, compares
the virial radius, $r_{\rm vir}$, to $r_s$ of the NFW profile, $c_{\rm vir}=r_{\rm
vir}/r_s$.

We specify the NFW model with the halo virial mass $M_{\rm vir}$ and
the concentration $c_{\rm vir}$ instead of $\rho_s$ and $r_s$. Here
the errors for these best-fit NFW parameters include the uncertainty
in the source redshift calibration for WL, $\overline{z}_s=1.0\pm
0.2$.  The typical halo concentration obtained by the four methods
described above is $c_{\rm vir}\simeq 7.5\pm0.5$, and the typical
virial mass is $M_{vir} \simeq1.2\pm 0.15\times
10^{15}M_{\odot}\,h^{-1}$ ($r_{\rm vir}\simeq1.84$ Mpc$\,h^{-1}$). The joint SL+WL fits yield $c_{\rm
vir}\simeq 7.15\pm0.5$ and $M_{vir} \simeq1.22\pm 0.15\times
10^{15}M_{\odot}\,h^{-1}$, agreeing with the recent result of Oguri et
al. (2009; $c_{\rm vir}\simeq 6.5^{+1.2}_{-0.7}$, $M_{vir}
\simeq1.05^{+0.28}_{-0.25}\times 10^{15}M_{\odot}\,h^{-1}$). This puts
Abell 1703 above the standard $c$--$M$ relation and manifests again the tension with the standard
$\Lambda$CDM model. This can be seen in Figure \ref{CM} where we plot confidence levels of the
concentration parameter derived for Abell 1703, along with $c$--$M$ relations including 1$\sigma$ uncertainties, deduced from simulations by Duffy et al. (2008) with WMAP5 parameters and scaled to $z_c=0.28$.
Since the concentration parameter depends on cluster formation time (as
discussed recently by Sadeh \& Rephaeli 2008) and though such conclusions have previously been reached for
several other massive clusters (e.g., Broadhurst et al. 2008, Umetsu et al. 2010), a similar comparison should naturally
be made for many other clusters in order to clearly establish that a statistically
significant trend is discerned.

\begin{table*}
\caption{Comparison of the
different NFW parameters obtained independently by different lensing
techniques. $\emph{Column 1:}$ The method; $\emph{Column 2:}$
Resulting concentration parameter, $c_{vir}$; $\emph{Column 3:}$
Resulting virial mass, $M_{vir}$, in $10^{15} M_{\odot}/h$;
$\emph{Column 4:}$ Reduced $\chi^{2}$ of the fit; $\emph{Column 5:}$
Q-value goodness of fit.}

\label{NFWfitTable}
\begin{center}
\begin{tabular}{|c|c|c|c|c|}
\hline
Method & $c_{vir}$ & $M_{vir}$ & Red. $\chi^{2}$ & $Q$-value\\
\hline
WL + Parametric SL &  $7.07\pm0.47$ &$1.18^{+0.12}_{-0.11}$& $0.33$& $0.99$\\
WL + Non-parametric SL &  $7.23\pm0.45$ &$1.26^{+0.13}_{-0.11}$& $0.43$& $0.99$\\
WL alone&  $7.71^{+2.22}_{-1.66}$ &$1.15^{+0.28}_{-0.22}$& $0.57$& $0.82$\\
WL + $\theta_e$ &  $7.92^{+1.34}_{-1.07}$ &$1.14^{+0.23}_{-0.21}$& $0.52$& $0.88$\\
\hline
\end{tabular}
\end{center}
\end{table*}

\section{Summary}

In this work we have produced an accurate full-range profile of the
cluster Abell 1703, based on combined WL and SL mass models applied to
exceptionally high quality space and ground based imaging data, including recent observations of the WFC3/IR on HST. This
profile adds to the relatively few clusters for which precise and
reliable mass profile have been constructed, including MS2137 (Gavazzi
et al. 2003, Merten et al. 2009), Abell 1689 (Broadhurst et al. 2005a,b, Lemze et al. 2008,
Umetsu \& Broadhurst 2008, Limousin et al. 2007), Abell 2218 (Kneib et
al. 1996, Abdelsalam, Saha \& Williams 1998) and Cl0024 (Zitrin et
al. 2009b, Umetsu et al. 2010).

We have applied two independent SL techniques in order to derive the
inner mass profile and to examine for consistency the basic assumption
of the parametric-model, that mass traces light. Both
techniques derive remarkably similar mass distributions with the same
major substructure present, and with similar overall radial mass profiles
(with a mean interior slope of $d\log \Sigma/d\log
\theta\simeq -0.5$). This inner profile matches well with
the WL data in the region of overlap, at around $\sim200$ kpc. The two SL
modelling methods cannot be distinguished in terms of the radial profile
or the overall large-scale mass distribution, but differ
mainly on fine scales principally owing to the inclusion of
cluster members in the parametric method.

Comparisons between other, different mass-modelling methods have been
made before (e.g., Valls-Gabaud et al. 2006, Coe et al. 2008, 2010,
Donnarumma et al. 2010, Meneghetti et al. 2010). Here we find that the
galaxy contribution is important to include in order to obtain
accurate predictive power, as is the case in the parametric model
(Zitrin et al. 2009b, Zitrin et al. 2010) which identifies and reproduces many multiple
images, even if initially constrained only by a few obvious
systems. On the other hand, being correlated to the initial light
distribution, the parametric model may at times be less flexible and therefore
over-sensitive to local luminous clumps or substructure, than the
non-parametric model which does not make any prior assumptions about the input distribution of mass (Liesenborgs et al. 2006).

We have made use of the reproduction ability of our parametric-model
and the remarkable multiply-lensed ring-like system next to the cD
galaxy, to uniquely constrain its projected mass $\emph{and profile}$ in the
inner region $\simeq1-5\arcsec$ ($\simeq4-22$ kpc). We have also found
that the low $M/L_B$ of $\sim6~(M/L)_{\odot}$ in this region can be
fully accounted for by stars, similar to cD galaxies in other
well-known clusters. The effect of baryons on halo density profiles is
still unclear and may be related to ``overcooling'' claimed in studies
of this effect (Barkana \& Loeb 2010, Duffy et al. 2010, Mead et al. 2010). Further
charactisation of cD galaxies from detailed lensing work will shed
more light on this still poorly understood class of objects.

\begin{figure}
 \begin{center}
   \includegraphics[width=80mm,trim=0mm 0mm 0mm 0mm,clip]{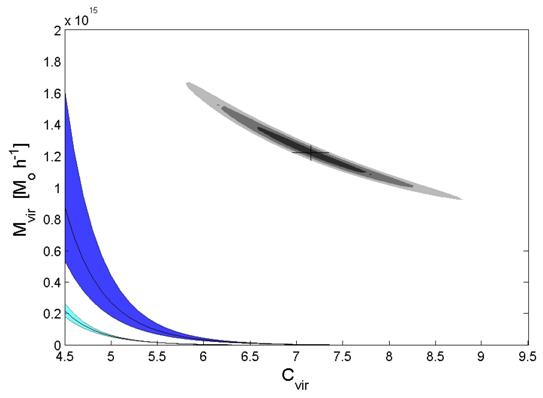}
 \end{center}
\caption{Chi-squared confidence levels (grayscale; 68.3\%, 95.4\%, and
99.7\% confidence levels) of the joint SL+WL data fits to an NFW
profile of Abell 1703, presented on the $c$--$M$ plane. Overplotted are the
expected $c$--$M$ relations and their 1$\sigma$ uncertainties,
presented in Duffy et al. (2008) with WMAP5 parameters, scaled to
$z_c=0.28$. The cyan shaded curve corresponds to the full Duffy et
al. (2008) sample, and the blue shaded curve corresponds to their
relaxed-halo sample. As can be seen Abell 1703 lies above the
standard $c$--$M$ relation.}
\label{CM}
\end{figure}

We have found that Abell 1703 lies above the standard $c$--$M$
relation (Figure \ref{CM}), similar to several other
well-known clusters for which detailed lensing-based mass profiles have
been constructed, adding to the claimed tension
with the standard $\Lambda$CDM model (e.g., Broadhurst et al. 2008, Umetsu et al.
2010; see also Sadeh \& Rephaeli 2008). Still, the overall level of systematic uncertainties may be
too large to allow a definite conclusion regarding a clear inconsistency with
$\Lambda$CDM predictions.
To further explore this apparent discrepancy a substantial multi-cycle Hubble
program has been established, based on an X-ray selected sample of relaxed
clusters so that no lensing bias is present in their selection (the \emph{CLASH}
program; see also \S1), for which the SL techniques applied here will be of great
value for deriving a statistically large and unbiased measurement of the
equilibrium mass profiles of galaxy clusters.

\section*{acknowledgments}

We thank Thomas Erben, the reviewer of this work, for many useful comments which improved this paper.
AZ acknowledges Eran Ofek and Salman Rogers for their publicly available Matlab
scripts. This research is being supported by the Israel Science Foundation grant 1400/10.
ACS was developed under NASA contract NAS 5-32865. This research is
based on observations provided in the Hubble Legacy Archive which
is a collaboration between the Space Telescope Science Institute
(STScI/NASA), the Space Telescope European Coordinating Facility
(ST-ECF/ESA) and the Canadian Astronomy Data Centre (CADC/NRC/CSA).
Part of this work is based on data collected at the Subaru
Telescope, which is operated by the National Astronomical
Society of Japan.

\bsp
\label{lastpage}

\end{document}